\documentclass{article}

\usepackage{ucs}
\usepackage[utf8x]{inputenc}
\usepackage[LGR,T2A,T1]{autofe}

\usepackage{graphicx}
\usepackage{amsmath}
\usepackage{url}

\usepackage{listings}
\begin{document}

\title{Exploiting the Cloud Control Plane for Fun and Profit}
\author{Josef Spillner\\
Zurich University of Applied Sciences, School of Engineering\\
Service Prototyping Lab (blog.zhaw.ch/icclab/)\\
8401 Winterthur, Switzerland\\
josef.spillner@zhaw.ch}

\maketitle

\begin{abstract}
Cloud providers typically charge for their services. There are diverse pricing models which often follow a pay-per-use paradigm.
The consumers' payments are expected to cover all cost which incurs to the provider for processing, storage, bandwidth,
data centre operation and engineering efforts, among others.
In contrast, the consumer management interfaces are free of charge as they are expected to cause only a minority
of the load compared to the actual computing services. With new service models and more complex and powerful
management abilities, it is time to rethink this decision. The paper shows how to exploit the control plane of AWS Lambda
to implement stateful services {\it practically for free} and under some circumstances even {\it guaranteed for free} which if widely
deployed would cause a monetary loss for the provider. It also elaborates on the consistency model for AWS Lambda.
\end{abstract}


\section{Motivation}
Cloud computing is an economic paradigm as much as it is a technical one. Its popularity can be attributed in part
to economic advantages such as a utility pricing model and spotmarkets \cite{cloudpricing,cloudmarkets}.
Applications on the cloud use a defined subset of programming interfaces (collectively a management or control API) to settle in,
connect to all dependency services and influence certain hosting parameters such as scheduling and scaling.
Beyond this use of the control plane, they rely heavily on the computing plane to transmit and store data, to invoke
compute services and to interact with other applications and services.

According to a widely used definition, the control plane facilitates the interaction between hosts (or other resources and services)
in the cloud infrastructure as well as between the client and the cloud \cite{selfservicecloudcontrolplane}.
As such, most cloud providers decided to not put a price tag on it. Given that conventionally no computing
can be performed on it, this has been a reasonable decision.

This way of thinking may now be over, however. With the Function-as-a-Service (FaaS) model, tiny stateless computing units
at nanoscale are promoted. The cost model is adapted accordingly to subsecond billing periods and micro- or nano-currency units. One FaaS offering is AWS Lambda for which
convincing cost benefits have been shown \cite{lambdacost}. A recent upgrade of Lambda also allows for tiny storage within
its control plane. While not accessible from the computing plane, this combination of free storage and
almost free computation will be explored and exploited in this paper for classes of stateful applications with
inherent access to the control plane.

After presenting helpful background information, the paper assesses the situation and then proceeds to
explain how to exploit it. During this process, it contributes a description of some of the characteristics
of AWS Lambda as well as a consistency model. Two control plane-bound applications will be introduced
and measured.
Finally, the findings will be discussed in the direction
of how future cloud applications and the underlying programmable platforms and infrastructure may be designed.


\section{Background}

\subsection{Cloud Services and Applications}

From an application perspective, the cloud computing service model is defined by an application
layer (SaaS) on top of two hosting layers (PaaS/IaaS). Platform and infrastructure services
primarily offer computing capabilities such as data transfer and queueing, virtual machine execution
and blob storage. Invoking these services may incur cost or may be for free, with or without
authentication in the form of tokens or other credentials. To the application and dedicated cloud management tools,
the platforms also offer a control plane which is bound to a user account and which generally only
allows authorised access to predefined management interfaces.

\begin{figure}[h]
\center
\includegraphics[width=.511\columnwidth]{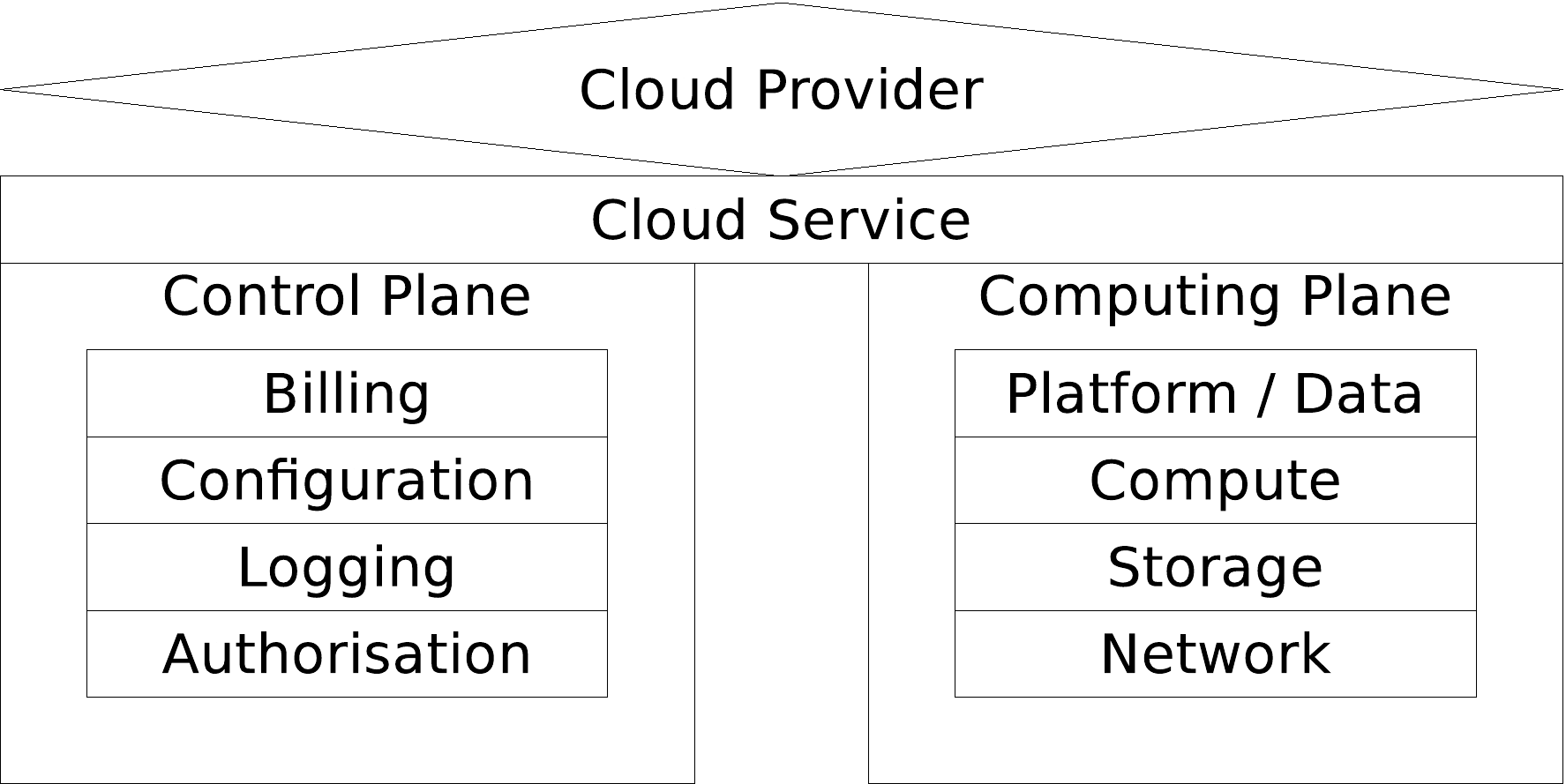}
\caption{Computing and control planes within a cloud service}
\label{fig:planes}
\end{figure}

The separation of the control plane from other planes originates in the networking domain
where other refers to the data (transmission) plane \cite{dataplane}. In the domain of cloud computing, the set of
others naturally extends to other resource types, namely compute and storage services, collectively
called the computing plane. Correspondingly, there are management or control interfaces which are
either provider-specific, as is the case with AWS, or generic, as for instance OCCI which
also includes application-level interfaces \cite{occi}.
The abstract notion of separated computing and control planes is shown in Fig. \ref{fig:planes}.

\subsection{Filesystems and Data Dispersal}

Cloud storage services on the computing plane offer large amounts of diverse storage areas
for blocks, files, blobs and archives. Examples on AWS include the Simple Storage Service (S3),
the Elastic Block Storage (EBS) and the Elastic File System (EFS) as well as Glacier
for long-term archiving. Glacier for instance stores a seemingly unlimited amount of files
of up to 40 TB each.
The cloud control plane typically limits the amount of data which can be stored to smaller amounts.
In order to achieve practical capacities, these small blocks need to be combined. This technique
resembles the organisation of storage devices where many fixed-size or variable-sized blocks are
combined and organised. The combination can be as simple as striping and as complex as optimal
erasure coding; and the organisation can be as simple as a table of linear contents and as complex
as a modern filesystem.

\begin{figure}[h]
\center
\includegraphics[width=.576\columnwidth]{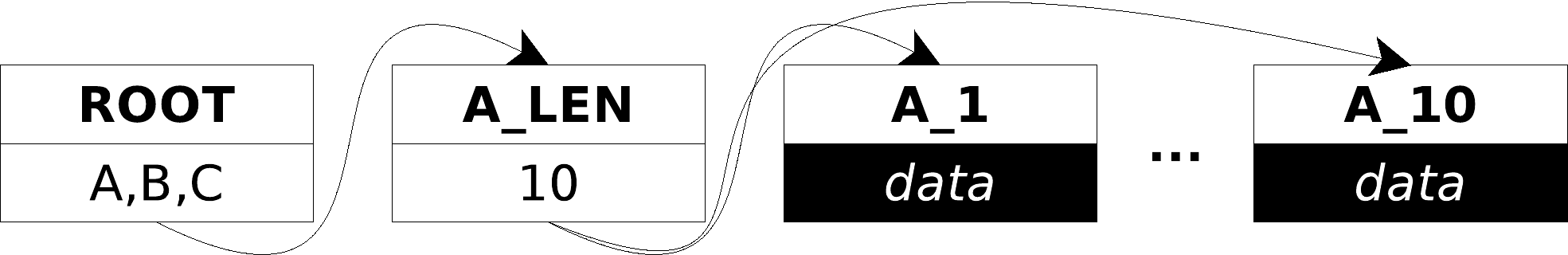}
\caption{Hierarchical table of contents structure}
\label{fig:toc}
\end{figure}

In this paper, we will make use of simple tree-shaped table-of-content (ToC) structures where a root block
contains pointers to chained subordinate blocks and collections of blocks are numbered to avoid an excessive
amount of pointers. The model maps well to environment variables in the form as shown in Fig. \ref{fig:toc}.

\section{Situation Analysis}

Our work focuses on AWS Lambda, a service offered since late 2014 by Amazon Web Services \cite{awssoftware}.
AWS Lambda is a service which hosts a set of Lambda functions per region. There are 16 regions
available although new ones are added every few months by the provider. Out of these, 14 are general-purpose
regions (exempting the US government cloud and the Chinese region) and 10 of those offer Lambda.
Each region is characterised by a human-readable name (e.g. US West / Northern California) and an
internal name which is also reflected in the domain name (e.g. \texttt{us-west-1}).

Each Lambda function consists of a name, a code implementation, a set of environment variables (since late 2016),
a trigger and further configuration.
The trigger could be an AWS API Gateway which translates external HTTP requests into Lambda invocations.
Seven other trigger types exist, most of which listen to events within other services of the same provider.
Upon triggered invocation, the code is executed and any data input and output is realised through services, for instance AWS S3 for
persistent blob storage. Despite the name, Lambda function code objects may encompass multiple actual functions
with private visibility.

\subsection{Function Code}

Lambda functions are implemented in a supported programming language and follow an expected interface
which requires at least one function, named entry point, to adhere to a certain signature. When invoked, a context object and
a JSON-formatted event structure is given as parameter, and a JSON structure is likewise expected as return
value. Supported programming languages include JavaScript in the Node.js flavour, Python, Java and C\#.
The entry point function in Python is defined as follows: \texttt{def lambda\_handler(event, context): return \{\}}.
The name of the function can be configured through the configuration setting $handler$. For brevity, we will use $f$ from now on.
The code implementation is mandatory and will be checked according to a basic parsing and code analysis upon deployment.
The minimal accepted, albeit invalid and not executable, implementation in Python would thus read as follows: \texttt{def f():pass}.
The net code size is then 12 bytes, even though each Lambda function adds an unavoidable overhead of 136 bytes presumably for configuration,
leading to a minimum possible size of 148 bytes for each Lambda function. It should be noted that the name of the Lambda
function itself does not influence this size; thus, a Lambda function verbosely and uniquely called $square\_root$ can call a
subordinate programming language function ambiguously called $f$.

The limits of AWS Lambda mandate a per-region maximum of 75 GB of deployed code. This implies that 544125924 or roughly 544 million
minimal Lambda functions could be deployed in each region. We assign this value to the constant $\#F$.

\subsection{Function Environment}

Lambda functions can be annotated with a set of key-value strings which are made available as environment variables
to function instances. While the same information could be represented in the function code, this model
cleanly separates the concern of implementation and configuration, and allows for faster updates in particular for
larger Lambda functions.
Each variable needs to adhere to the following syntactical regular expression: $[a-zA-Z]([a-zA-Z0-9\_])+]$.
This implies that a two-character name (e.g. $AA$) is the shortest possible one.
Furthermore, each value needs to adhere to the expression $[^\wedge,]*$ which implies that commas are not allowed.

The limits of AWS Lambda concerning the environment mandate a maximum size of 4096 bytes per Lambda function.
This limit is enforced over the JSON representation of the set of variables which reads as follows:
$\{"AA":"..."\}$. This implies a loss of 9 bytes so that 4087 bytes remain usable for the value in the extreme
case of only having one variable. We assign this value to the constant $S$.
For completeness, it should be mentioned that further environment variables are provided by AWS Lambda itself
which do not count into the quota, for instance $AWS\_SESSION\_TOKEN$ or $AWS\_LAMBDA\_-FUNCTION\_VERSION$.

\subsection{Combination}

The environment variables configuration of AWS Lambda can be used to store arbitrary binary data. In this case,
it would have to be represented as chunks in a suitable encoding. A safe encoding resulting in only uppercase and lowercase characters,
numerals and equal signs is Base64 which reduces the usable net size by 25\% of $S$. Due to padding on multiples of
three bytes of input or four bytes of output respectively, this means that 3063 arbitrary bytes can be stored. We assign this value to the constant $B$, i.e. $B = 3 \lfloor \frac{S}{4} \rfloor$.

\begin{figure}[h]
\center
\includegraphics[width=.463\columnwidth]{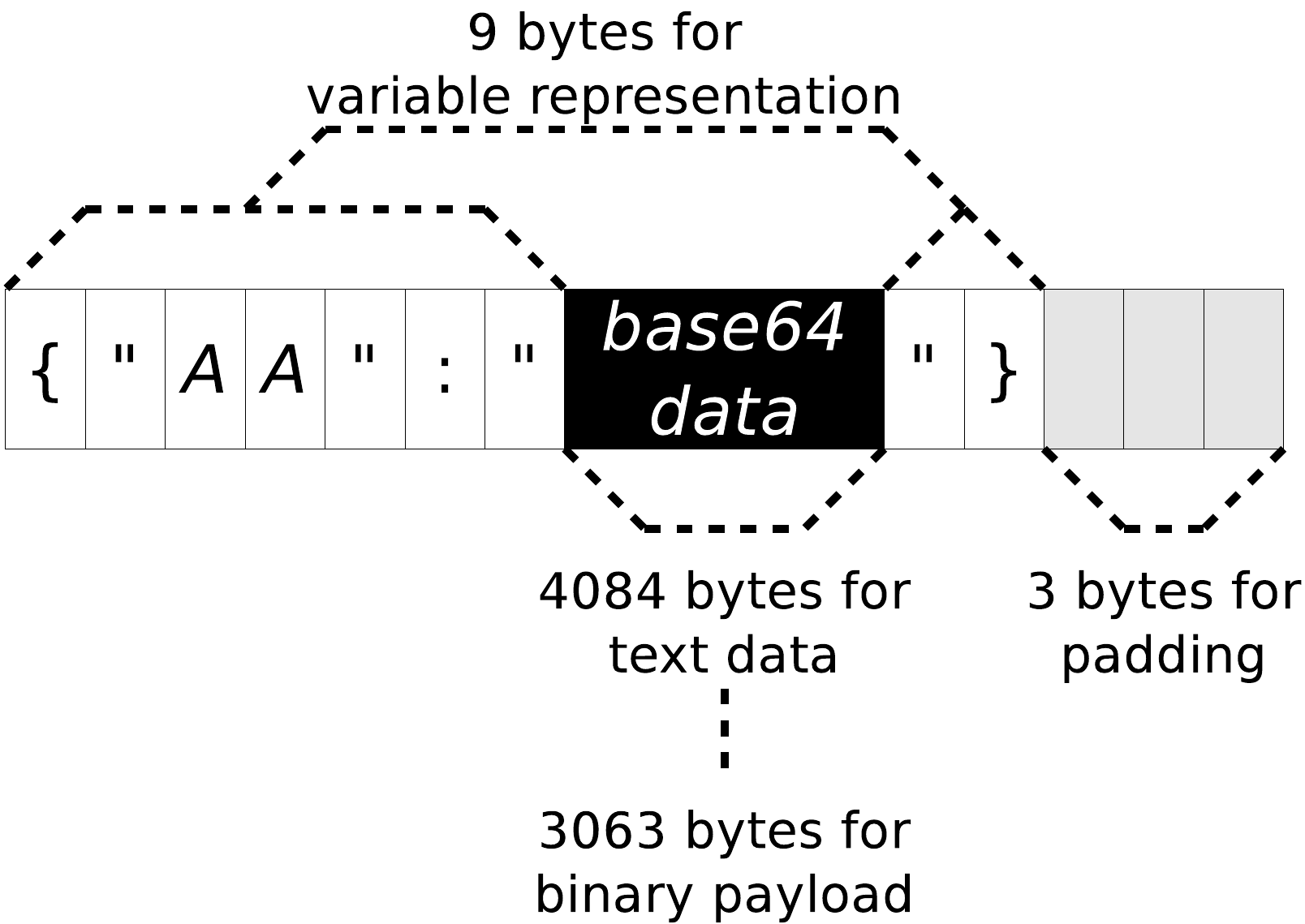}
\caption{Byte-level representation of environment variables configuration in AWS Lambda}
\label{fig:bytearray}
\end{figure}

Fig. \ref{fig:bytearray} summarises the allocation of 4096 bytes when the goal is the maximisation of the payload area for binary data.

Eq. \ref{eq:combination} combines our analytical results on the function code and on the respective environment.

\begin{align}
\label{eq:combination}
 \#F &= 544125924 \nonumber \\
   B &= 3063 \; bytes \nonumber \\
\#FB &= 1666657705212 \; bytes \nonumber \\
     &\approx 1.51 \; TB
\end{align}

Hence, we conclude that in total about 1.51 TB of data can be stored per region in the unaccounted control plane
of AWS Lambda which corresponds to a typical notebook solid state disk capacity. Across all regions, this value
is multiplied to 15.1 TB.
Through compression, the amount of data stored could be further increased.

When the function should be invocable, its code needs to be adapted. A minimal implementation is $f_1$
in Listing \ref{lst:retrieval} which increases the code size by 31 bytes. When the
data should furthermore be globally accessible, for instance without authentication, the AWS API Gateway
needs to be switched on with a dedicated resource in front of each Lambda function. The code then
grows by another 39 bytes due to the necessary JSON format of the return value as implemented
in $f_2$ in the same listing.

\begin{lstlisting}[caption=Lambda environment variable data retrieval,label=lst:retrieval,mathescape]
import os
def f$_1$(e,c):return os.getenv("AA")
def f$_2$(e,c):return {"statusCode":200,"headers":{},"body":os.getenv("AA")}
\end{lstlisting}

The function then becomes accessible by POST requests to the resource
\texttt{https://<api-id>.execute-api.<region>.amazonaws.com/prod/<lambda-\\ name>}.
One important limitation of the gateway is that only 3 (fixed, not increasable) $CreateDeployment$ requests per minute are permitted
which contradicts the rapid instantiation of micro- or nanoservices requiring dozens or more resources in a short
period of time. Likewise, the number of resources is limited. Not considering these limitations,
the increased function size reduces the theoretic maximum of retrievable data to 1.25 TB and 1.03 TB,
respectively.


\section{Situation Exploitation}

We argue that by placing computational elements into the control plane of a cloud service, the
provider's pricing model can be effectively forgone. Our contribution to demonstrate this hypothesis
are two distinct software applications both of which benefit from the unmetered network connection
to the control plane.
The first application is a backup tool for storing and retrieving data which furthermore exploits the storage capabilities to achieve a {\it guaranteed for free}
service level. The second application is a special-purpose database which exploits both storage and compute capabilities
and due to the pricing model achieves at least a {\it practically for free} service level.
Both applications are provided as open source prototypes.

\subsection{Backup Application}

Lambackup is a shell script which splits files into chunks to store them into the environment variables section of generated Lambda functions.
Furthermore, the script can retrieve and reassemble the files. The length of a file can be determined deterministically by storing one more
chunk of information containing the file length on position zero (LEN), or heuristically by retrieving chunks until an error occurs.
The script can also maintain a growing list of filename records including a counter chunk as table of contents (TOC).

The experiments are conducted using two reference files, the Apache 2.0 licence text (11 kB, called \textit{small} in the following) and a binary PDF
of the Jolie paper from ACM SAC'13 (142 kB, called \textit{large}) \cite{jolie}.

\begin{figure}[h]
\center
\includegraphics[width=.8\columnwidth]{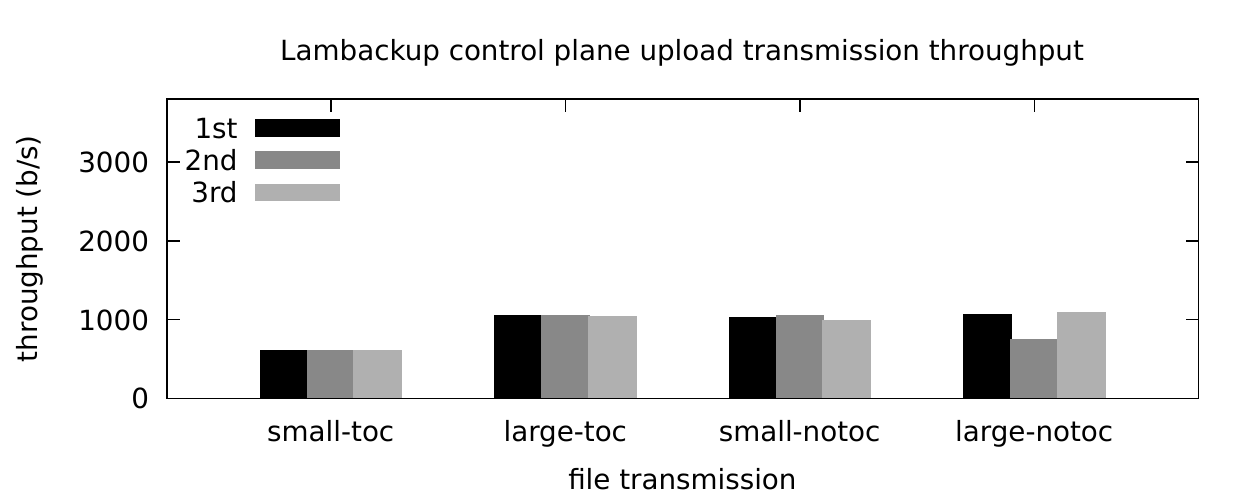}
\caption{Lambackup backup file transmission throughput with and without recording of file length and table of contents; 3 consecutive runs}
\label{fig:logbackup}
\end{figure}

Fig. \ref{fig:logbackup} gives an impression of the transmission performance of Lambackup when storing the two files which take 4 and 48 functions for storage, respectively,
one time in minimal configuration and another time in the default LEN+TOC configuration. It is evident that the throughput is extremely low
but stabilises for files larger than just a few kilobytes to a predictable $1 kB/s$. The overhead of using LEN+TOC becomes then negligible.

\begin{figure}[h]
\center
\includegraphics[width=.8\columnwidth]{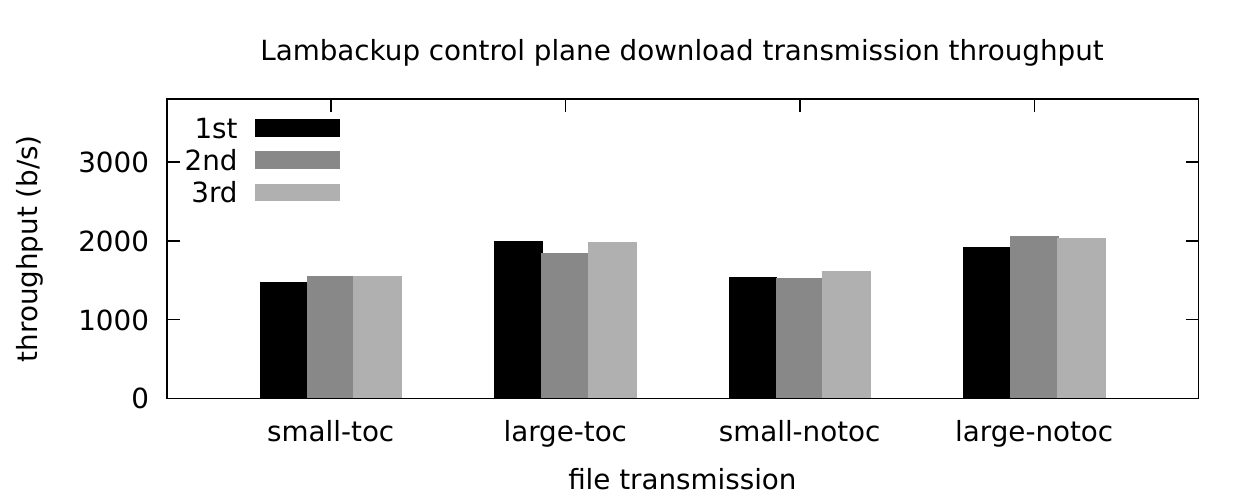}
\caption{Lambackup restore file transmission throughput associated to Fig. \ref{fig:logbackup}}
\label{fig:logrestore}
\end{figure}

The restore operation is faster as no Lambda functions need to be created and no overhead due to transmitting the function code itself occur. Fig. \ref{fig:logrestore} demonstrates the corresponding restore throughput. The reachable practical limit is about $2 kB/s$.

\begin{figure}[h]
\center
\includegraphics[width=.8\columnwidth]{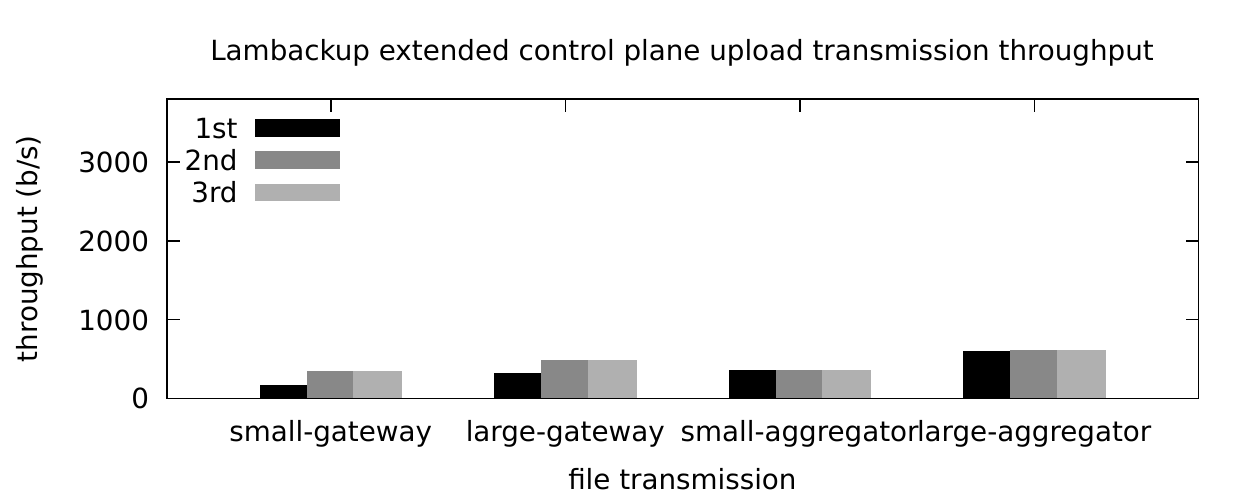}
\caption{Lambackup fast backup file transmission throughput; 3 consecutive runs}
\label{fig:logbackupfast}
\end{figure}

\begin{figure}[h]
\center
\includegraphics[width=.8\columnwidth]{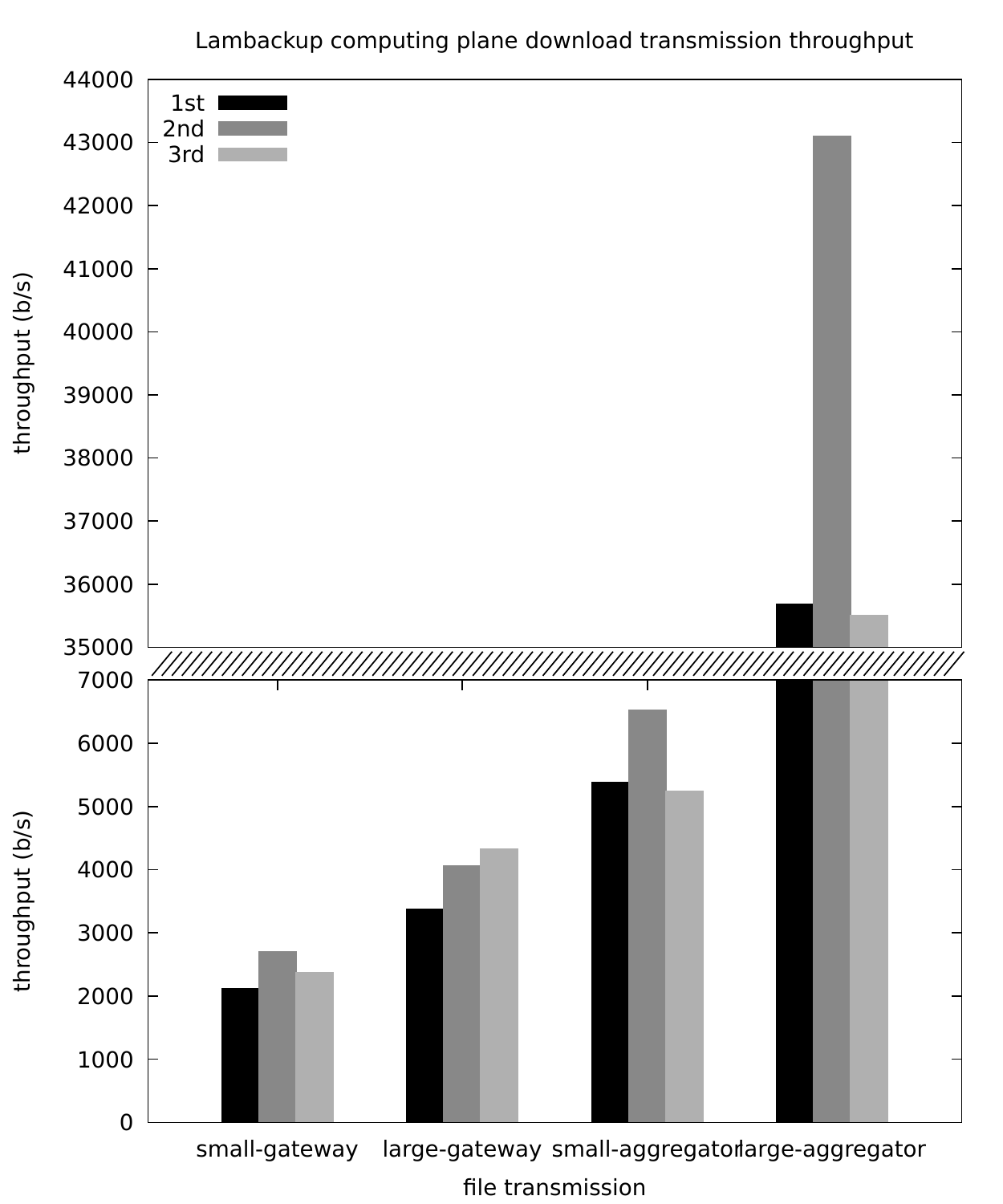}
\caption{Lambackup fast restore file transmission throughput associated to Fig. \ref{fig:logbackupfast}}
\label{fig:logrestorefast}
\end{figure}

Another option FST exists to speed up the retrieval at the expense of involving the computing plane, leading to only practically
free data retrieval in constrast to the guaranteed free transmission in both directions.
The download may become fast by just invoking a Lambda function per chunk which returns the data as shown before, or invoking
a aggregation single function which assembles the entire file before returning it by internally calling the chunk functions,
or by connecting an API Gateway as frontend service. The first of these options does not accelerate the download,
partially invalidating this hypothesis, but the other two do.
Figs. \ref{fig:logbackupfast} and \ref{fig:logrestorefast} show the decreased backup throughput but in turn increased restore throughput when applying the FST options for downloading chunks through the API Gateway and through an aggregation function invoked via the control plane.
Compared to Figs. \ref{fig:logbackup} and \ref{fig:logrestore}, these invocations also use LEN+TOC.
As Lambda by default interrupts functions after a duration of $3s$, the aggregation function needs to be configured with a much higher
limit to avoid the interruption which would otherwise affect even small files, up to the maximum of $300s$.

Listing \ref{lst:aggregator} shows the implementation of the aggregation function.
It iterates heuristically over the chunks following the naming scheme of the split command
which names chunks $dataaa, ..., datayz, dataaaaa, ...$.

\begin{lstlisting}[caption=Lambda environment variable data aggregator,label=lst:aggregator,mathescape,float=*]
import json, os
from boto3 import client as bc
lambda = bc('lambda')
def f(e, c):
  comb = ["a", "a"]
  responses = []
  fn = os.getenv("functionname")
  while True:
    try:
      response = lambda.invoke(FunctionName=fn + "data" + "".join(comb), Payload="{}")
    except:
      break
    response = json.loads(response["Payload"].read())
    #access ["body"] when using API Gateway
    responses.append(response)
    if comb[1] == "z":
      comb[1] = "a"
      comb[0] = chr(ord(comb[1]) + 1)
    else:
      comb[1] = chr(ord(comb[1]) + 1)
  return "".join(responses)
\end{lstlisting}

Table \ref{tab:comparison} summarises the results and compares the characteristics of the environment data transmission options.
There is a clear trade-off between the increased download performance gains and the increased cost, tighter constraints and decreased
upload performance of the two options which use the compute services.

\begin{table}[htb]
\centering
\caption{Comparison of Lambda environment data transmission; up- and download in b/s.\label{tab:comparison}}
\begin{tabular}{|l|l|l|l|l|} \hline
\textbf{Option}		& \textbf{Upload}& \textbf{Download}	& \textbf{Cost}	& \textbf{Limits}	\\ \hline

Default	storage		& 939--1016	& 1608--2023		& 0		& --			\\ \hline
LEN+TOC storage		& 604--1043	& 1541--1975		& 0		& --			\\ \hline
Lambda retrieval	& - as above -	& 1415--1724		& $\approx 0^{*}$& 6 MB payload		\\ \hline
API Gateway		& 336--481	& 2366--4324		& $\approx 0^{**}$& 3 req/min		\\ \hline
Aggregator		& 349--608	& 5234--35507		& $\approx 0^{*}$& 3--300 s duration	\\ \hline
\multicolumn{5}{l}{*: Lambda pricing: 0.20 µUSD per call and 16.67 µUSD per GB/s beyond}		\\
\multicolumn{5}{l}{free tier; **: additionally API Gateway pricing: 3.50 µUSD per call}			\\
\multicolumn{5}{l}{and 0.09 USD per GB traffic beyond free tier}					\\ \hline
\end{tabular}
\end{table}

To illustrate this trade-off, consider a file with a size of 1 GB. It would need to be split into 350553 chunks
encoded into environment variables.
The upload would take around 284 hours, making it impractical for anything but delay-tolerant background jobs.
The download could be accelerated to finish in less than 9 hours by invoking a series of an estimated 6000 aggregator instances
at the cost of still 0 USD within the free tier or 0.07 USD beyond it.
The combination of API Gateway and aggregation function would maximise the download throughput but also the cost.
For comparison, storing the same amount of data into S3 for one year would cost 0.27 USD beyond the free tier
in addition to the transmission cost of USD 0.004 for each 10000 requests.

Adding compression, the file size can be reduced significantly. The ratio is 26.9\% with the recent
Brotli algorithm and still 44.6\% with the Gipfeli algorithm for the text file, or 3056 of 11358 bytes. For the binary file,
the ratios are 99.0\% and 99.5\% respectively. Other compressors such as bzip2 or xz achieve varying results in between the
two mentioned ones.

The conclusion of these experiments is that depending on the time available for a storage and retrieve task, the AWS Lambda control
plane offers a suitable concealment for smaller critical data which will not cause cost over longer periods of time.
Potential use cases are dead drops and political leaks applications, high-value file and digital key vaults and
personal file shares.

\subsection{Database Application}

LaMa is a pseudo-relational database management system consisting of a client tool and a set of Lambda functions.
As unique design criterion, all tables, columns and data as content of the columns are stored in the environment
variables section of Lambda's configuration.
Only read-only queries are performed through Lambda functions whereas all modifications are required
to use the configuration interface due to the service's security model which does not allow access to the
control plane from the computing plane.

LaMa's data model is dynamic, supporting only one data type ($DYNAMIC$) with a basic subset of the Structured
Query Language (SQL). Of interest in such a design are two properties, performance and correctness.
The implementation, whose query frontend is derived from StealthDB \cite{stealthdb}, has been measured
by running a sequence of SQL commands for 100 times. The sequence
drops the table if it exists, (re-)creates it as single-column table, inserts 100 values and queries
the entire data without $WHERE$ clause.

\begin{figure}[h]
\center
\includegraphics[width=.8\columnwidth]{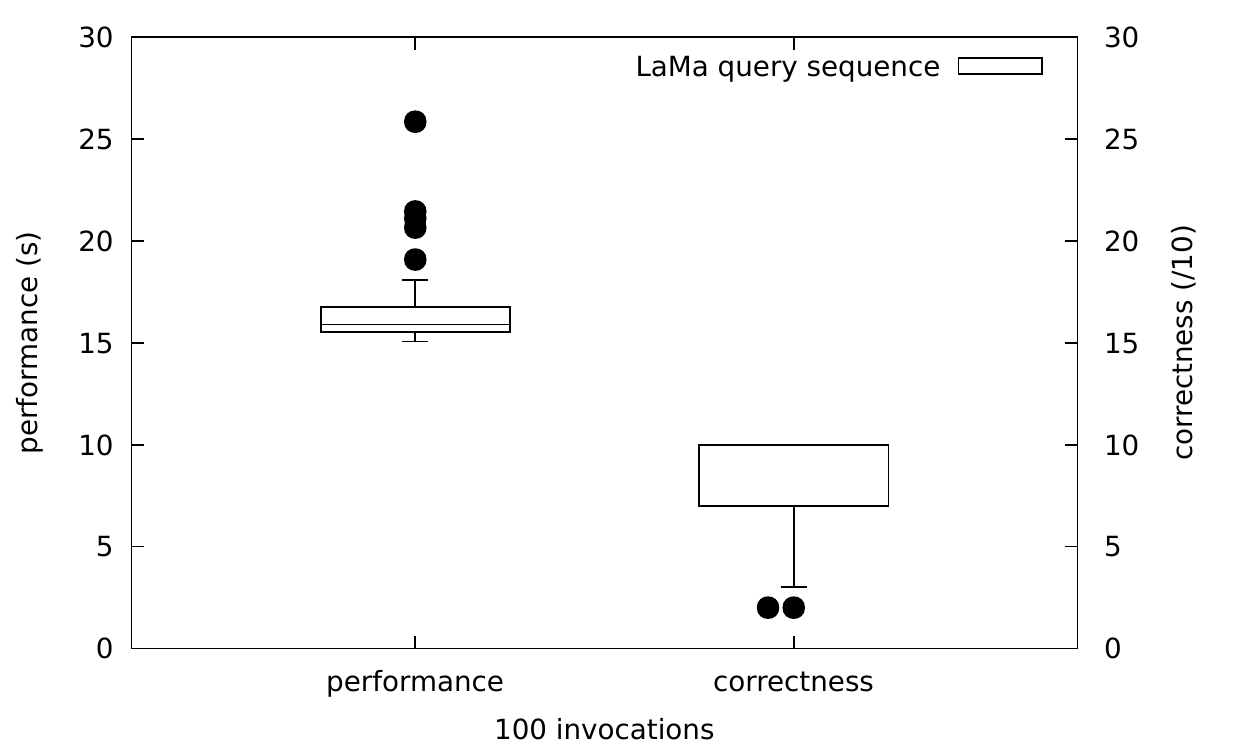}
\caption{Benchmark for performance and correctness of LaMa}
\label{fig:bench}
\end{figure}

Fig. \ref{fig:bench} shows the results. The execution performance is relatively stable around an average of
$16.4s$ on the test system with a few slower outliers having up to 58\% higher time requirements.
More interesting is the deviation in the number of query results which in about one third of all cases
ranges from 2 to 9 instead of the expected 10.

\begin{figure}[h]
\center
\includegraphics[width=.8\columnwidth]{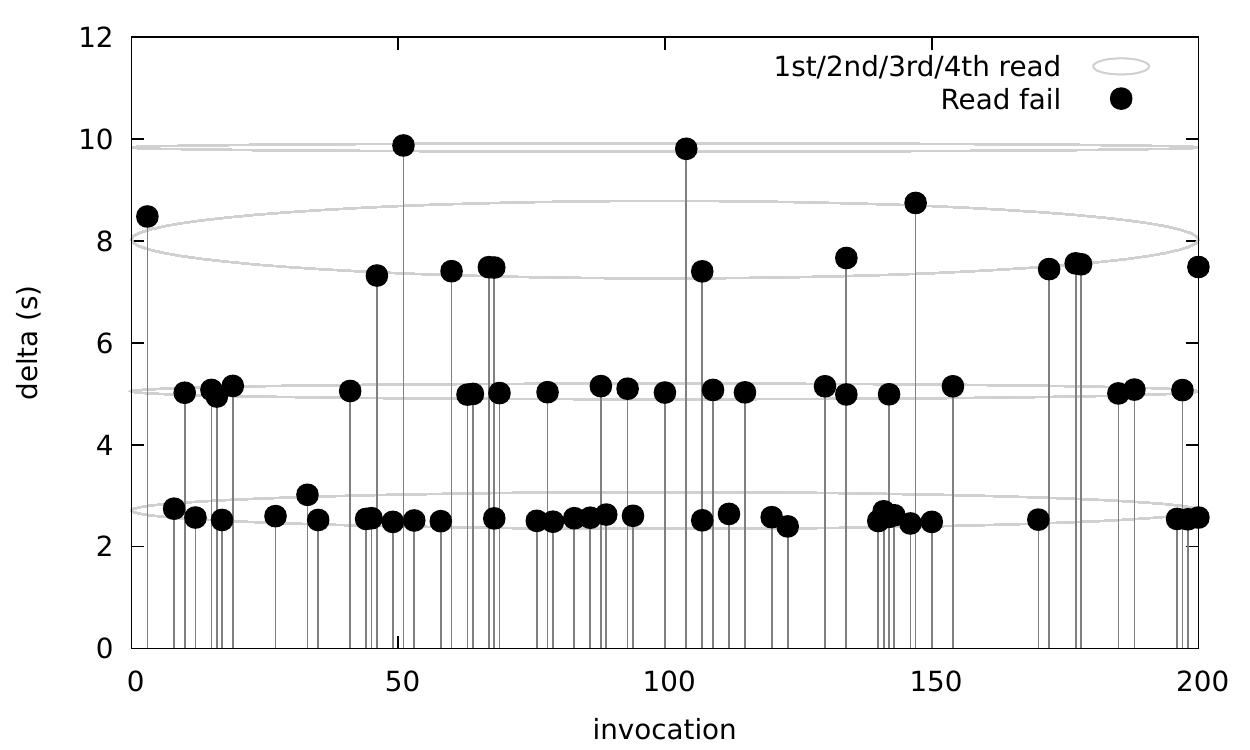}
\caption{Eventual consistency results}
\label{fig:inconsistency}
\end{figure}

The limited correctness can be explained by another, more focused experiment
which runs 200 rounds of random environment variable assignments, each followed
by 10 consecutive reads. In total, 67 out of 2000 reads fail, about half of them in the
first read, about one third in the second, and the remainder in the third and fourth.
Five failures appear twice with or without a successful read in between.
No error occurred after the 4th read attempt which coincides with a maximum
error interval of $9.87s$. We conjecture that the internal synchronisation of Lambda's
configuration is set to $10s$. Indeed, a second run of the same experiment with
a $10s$ delay after the write yielded no read failures which helps application authors
in making their applications less eventual and more consistent.
Fig. \ref{fig:inconsistency} contains the consistency measurement results in a graph
with clustered read failures. The execution performance stability despite network jitter is remarkable
and leads to non-overlapping clusters.

\begin{figure}[h]
\center
\includegraphics[width=.511\columnwidth]{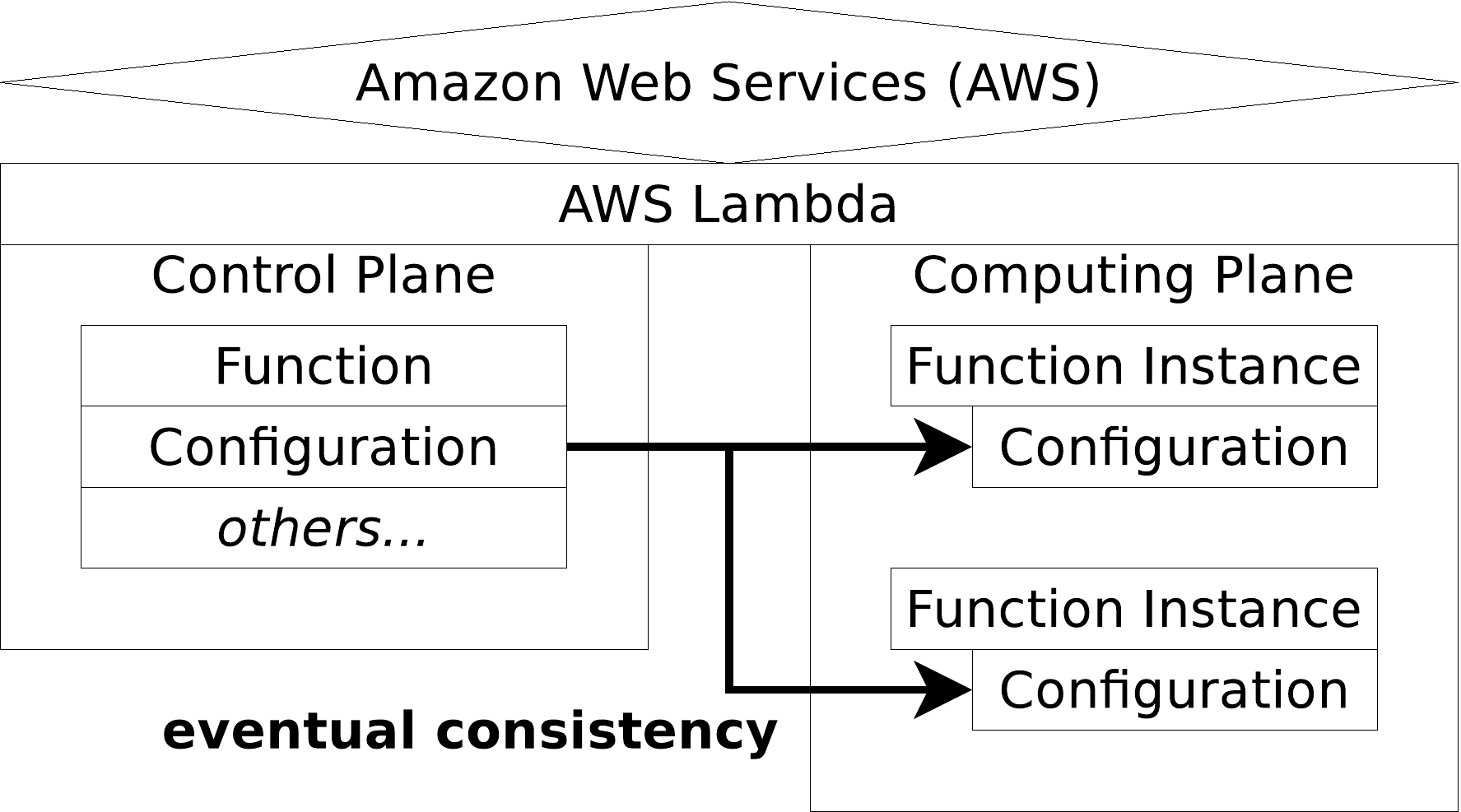}
\caption{Lambda-specific computing and control planes}
\label{fig:awsplanes}
\end{figure}

The presence of eventual consistency and its boundary characteristics are thus far not documented for AWS Lambda.
Other services of the same provider are described more completely, for instance EC2 for whose control plane
the eventual consistency is mentioned, albeit without details about boundary conditions.
We argue that expressive service descriptions would help to make such characteristics explicit.
In the absence of a clear documentation, we assume that Lambda instances have lazy-synchronised
configuration copies. Fig. \ref{fig:awsplanes} refines the introductory figure adapted to how
AWS Lambda is known to work.

The conclusion of these experiments is that several applications, including append-only emitters of
small data portions such as sensors, can exploit the AWS Lambda control plane, along with eventual consistency-aware
recipients of the raw or aggregated data.


\section{Discussion}

It appears unusual to massage a cloud service's control plane into a useful service in itself upon which
meaningful applications are built. Whether this becomes a sound technique or remains a singular
proof of concept needs to be answered by future work. Concerning AWS Lambda, both the advantages
(zero cost) and disadvantages (severe limits and hardly predictable eventual consistency) of the control
interface are now palpable for future system designs.

Stateful Lambda functions can only be implemented with explicit access to the control plane.
The service model is therefore not suitable for holding state per function in the general case or per function instance
in any case. Among other effects, this limitation affects ongoing research work to transform object-oriented
code to functional code which requires stateful instance attributes.

The requirements for future programmable platforms and infrastructure include better discoverability
(through manuals and processable descriptions), better composability and programmability (refer to the non-trivial integration
between AWS Lambda and API Gateway) and better procedures. A suitable model would be working on
a disconnected and versioned configuration which is model-checked and then enacted or rolled back during a single
controllable transaction. The role of the control plane would then change considerably.

Returning to the opening remarks about cloud computing being an economical model, it becomes apparent
that for instance control plane-based database functions would need to be priced differently depending
on whether a function call is involved or merely a CRUD operation. Ultimately, performance drawbacks
are not always a convincing argument against services which work {\it guaranteed for free} or at least
{\it practically for free}. The argument is weakened by the fact that most cloud storage services
are already very low-cost and thus any difference would only be noticeable at larger scale.


\section*{Repeatability}

The implementations of Lambackup and LaMa as well as the scripts used in the referred experiments and
the obtained reference results are made publicly available. We encourage the creative use of these artefacts
for repeatability of the results and for confirmability of the presented findings.
The corresponding repository is \url{https://github.com/serviceprototypinglab/lambdacontrolplane}.

\section*{Acknowledgements}

This research has been supported by an AWS in Education Research Grant which helped us to run our experiments on AWS Lambda as representative public commercial FaaS.

\bibliographystyle{unsrt}
\bibliography{cloudcontrolplane}

\end{document}